\documentclass[a4paper,11pt]{article}
\usepackage[utf8]{inputenc}
\usepackage[english]{babel}
\usepackage{braket}
\usepackage{xspace}
\usepackage{bbm}
\usepackage{mathdots}
\usepackage{stackrel}
\usepackage[dvipsnames]{xcolor}
\usepackage{appendix}
\usepackage{hyperref}
\hypersetup{
 colorlinks=true,
 linkcolor=blue,
 anchorcolor = blue,
 citecolor = blue,
 filecolor = blue,
 urlcolor = blue
}
\usepackage{mathtools}
\usepackage{bbm}
\usepackage{comment}
\usepackage{subfigure}
\usepackage[T1]{fontenc}               
\usepackage[left=3cm,
            right=2.5cm,
            top=2.5cm,
            bottom=3cm
            ]{geometry}                
\usepackage{graphicx}
\usepackage{mathrsfs}
\usepackage{appendix}
\usepackage{fancyhdr}
\usepackage{amsmath,                   
            amssymb,                   
            amsthm}                    

\usepackage{setspace}
\usepackage{cite}
\usepackage{cancel}
\usepackage{bm}
\usepackage[margin=30pt, bf, font=small, center, justification=justified]{caption}[2004/07/16]

\usepackage{setspace} 

\usepackage{tikz}

\title{\bf Inhomogeneous quenches and GHD in the $\nu = 1$ QSSEP model}

\author{Angelo Russotto$^1$, Filiberto Ares$^1$, Pasquale Calabrese$^{1}$, Vincenzo Alba$^2$}

\date{}

\begin{document} 

\maketitle

{\small
\vspace{-5mm}  \ \\
{$^{1}$}  SISSA and INFN Sezione di Trieste, via Bonomea 265, 34136 Trieste, Italy\\[-0.1cm]
\medskip
{$^{2}$}   Dipartimento di Fisica and INFN Sezione di Pisa, Largo Bruno Pontecorvo 3, 56126  Pisa, Italy\\[-0.1cm]
\medskip
}

\begin{abstract}
We investigate the dynamics of the $\nu=1$ Quantum Symmetric Simple Exclusion Process starting from spatially inhomogeneous initial states. This one-dimensional system of free fermions has time-dependent stochastic hopping amplitudes that are uniform in space. We focus on two paradigmatic setups: domain-wall melting and the expansion of a trapped gas. Both are investigated by extending the framework of quantum generalized hydrodynamics to account for the underlying stochastic dynamics. We derive the evolution of the local quasiparticle occupation function, which characterizes the system at large space–time scales, and analyze the resulting entanglement spreading. By incorporating quantum fluctuations of the occupation function and employing conformal field theory techniques, we obtain the exact contribution to the entanglement entropy for each individual noise realization. Averaging over these realizations then yields the full entanglement statistics in the hydrodynamic regime. Our theoretical predictions are confirmed by exact numerical calculations. The results presented here constitute the first application of quantum generalized hydrodynamics to stochastic quantum systems, demonstrating that this framework can be successfully extended beyond purely unitary dynamics to include stochastic effects.
\end{abstract}

\tableofcontents

\section{Introduction}

Over the past decade, our understanding of non-equilibrium many-body quantum systems has advanced considerably. This progress has been driven both by the development of novel theoretical tools and by the emergence of experimental platforms capable of simulating these complex systems. In particular, the formulation of generalized hydrodynamics (GHD)~\cite{bcdf-16, cdy-16} (see~\cite{bastianello-22, d-20,e-rev,doyon-25} for recent reviews) has opened the way to the systematic study of quenches in integrable systems with inhomogeneous initial conditions. Typical examples include situations in which the two sides of a system are initially prepared in different configurations and subsequently connected at a junction, or cases where the confining potential of a gas is suddenly released, allowing it to expand into a larger region~\cite{abfpr-21}. The main idea behind GHD is that, at large space–time scales, inhomogeneous quantum systems can be locally described in terms of a quasiparticle momentum occupation function. The evolution of this occupation function obeys a continuity equation, with the quasiparticles moving ballistically at effective velocities. GHD readily yields results for conserved charges and currents, which have been validated across different models, see e.g.~\cite{pncbf-17, bvkm-17, ddky-17, clv-18, dyc-18, bac-19, cddky-19}, and cold-atom experiments~\cite{sbdd-19, malvania-21, horvath-25,dubois-26}. Although GHD does not directly capture intrinsically quantum features, such as the spreading of entanglement and quantum correlations, these effects can nevertheless be incorporated by re-quantizing it in terms of an inhomogeneous Luttinger liquid theory~\cite{rbd-19, rcdd-21,srcd-23,tscvd-24} (see also \cite{f-17} for an alternative approach).

While GHD primarily applies to integrable models, we here extend its methods to inhomogeneous quenches in a stochastic many-body quantum system: the $\nu=1$ Quantum Symmetric Simple Exclusion Process (QSSEP). This model is the simplest instance of a larger family of free-fermion systems, the $\nu$-QSSEP~\cite{alba-25}, characterized by hopping amplitudes between neighboring sites that are random in time and periodic in space, being equal every $\nu$ sites. In the case $\nu=1$, the noise is spatially homogeneous, whereas in the opposite limit $\nu\to\infty$ it becomes fully inhomogeneous, corresponding to the original QSSEP~\cite{bbj-17, bbj2-19}. The latter has emerged as a paradigmatic model for the quantitative study of quantum diffusive dynamics in the route toward a quantum version of macroscopic fluctuation theory~\cite{mft,bernard-21}, with its average dynamics matching its classical counterpart, the SSEP~\cite{mallick-15}. The fluctuations of the stationary state and its entanglement entropy have been characterized in the QSSEP~\cite{bb-25, bbj3-20, hb-23, bp-21}, as well as in several of its generalization~\cite{bj-19, bj-21, jkb-20, bjsw-25, abjsw-26, crdl-25, sbn-25}. Their dynamics has been investigated under various protocols, including bipartite initial states~\cite{behm-22}. Nevertheless, a comprehensive understanding of the time evolution of the entanglement entropy is still lacking.

By contrast, in the $\nu=1$ QSSEP,  the presence of spatially homogeneous noise enables a modified quasiparticle description of entanglement spreading from translationally invariant initial states~\cite{alba-25}, closely analogous to the quasiparticle picture developed in integrable systems~\cite{cc-05, fc-08, ac-17, ac-18,c-20}. In this stochastic setting, quasiparticles perform Brownian motion in each noise realization, leading to a diffusive growth of the entanglement entropy, rather than the ballistic propagation and linear growth characteristic of integrable dynamics. Within this stochastic quasiparticle framework, it is possible to compute the time evolution of arbitrary functions of the reduced density matrix of a subsystem, including, for example, the exact full counting statistics of entanglement-related observables~\cite{raca-25}.
These findings suggest that a generalized version of GHD, in which quasiparticles undergo effective Brownian motion,  provides a natural and powerful framework for describing inhomogeneous quenches in the $\nu = 1$ QSSEP. Motivated by this insight, we focus here on two paradigmatic and well-studied setups: the domain-wall melting and the expansion of a confined gas of free fermions at half-filling. Both scenarios have been extensively analyzed in the non-random counterpart of the $\nu=1$ QSSEP, namely the standard tight-binding model~\cite{antal-99, karevski-02, ogata-02, pk-07, antal-08, lancaster-10, allegra-16, viti-16, vicari-12, ah-14, gruber-19, hrs-04, eisler-25, ep-14, er-13}. In particular, the exact evolution of the entanglement entropy in these settings has been obtained using quantum GHD~\cite{dsvc-17, skcd-21}. 
These inhomogeneous quenches therefore provide an ideal testing ground for assessing the impact of stochastic dynamics and for benchmarking the generalized hydrodynamic description.
Building on and adapting these techniques, we derive analogous results for the $\nu=1$ QSSEP.

The paper is organized as follows. In Sec.~\ref{sec:model}, we introduce the $\nu = 1$ QSSEP model and the two protocols analyzed in this work. In Sec.~\ref{sec:dwsec}, after reviewing the main results of QGHD needed for our analysis, we study the first protocol: the domain-wall melting. In Sec.~\ref{sec:freexpsec} we investigate the free expansion setup. Finally, in Sec.~\ref{sec:conclusions} we draw our conclusions. We also include three appendices. In Appendix~\ref{app:deriv}, we derive the stochastic differential equation for the local occupation function. In Appendix~\ref{app:freeexp}, we discuss the free expansion of a gas when the hopping amplitudes are non-random, but complex, which is helpful to understand the free expansion in the $\nu=1$ QSSEP. In Appendix~\ref{app:num_tech}, we detail the numerical methods employed to check our results. 

\section{Quench protocols}\label{sec:model}
In this section, we introduce the two quench protocols analyzed in this work: the domain wall melting (that will be studied in Sec.~\ref{sec:dwsec}) and the free expansion of a confined gas (resp. Sec.~\ref{sec:freexpsec}). 
In both cases, we consider a one-dimensional system of non-interacting spinless fermions whose unitary dynamics is governed by random complex hopping amplitudes that are homogeneous in space but stochastic in time. This corresponds to the $\nu = 1$ QSSEP model introduced in Ref.~\cite{alba-25}. The time evolution is governed by the following stochastic Hamiltonian generator
\begin{equation}
\label{eq:model_dH}
    d H_t = \sqrt{D} \sum_{j} \left( dW_t c^{\dagger}_j c_{j+1} + h.c.\right),
\end{equation}
where $c_j$ ($c^{\dagger}_j$) are the annihilation (creation) operators of a spinless fermion at the site $j$. The hopping amplitude $dW_t$ is a complex It\^o increment that satisfies $dW_td W_t =\overline{dW}_t\overline{d W_t}= 0$ and $dW_t\overline{d W_t} = dt$. The constant $D$ only determines the size of the fluctuations of the noise.
The state of the system is described at any time $t$ by the density matrix $\rho_t$, which evolves as 
\begin{equation}\label{eq:time_ev}
\rho_{t+dt} = e^{-i dH_t} \rho_t e^{i dH_t}.
\end{equation}
The main difference between the two quench protocols is the initial state of the system. Let us now discuss the two main protocols of interest. 

\textbf{Domain wall melting (Sec.~\ref{sec:dwsec}).}  In this case, the system is prepared at time $t = 0$ in the product state
\begin{equation}
\label{eq:DWdef}
    \ket{\mathrm{DW}} = \prod_{j<0}c^{\dagger}_j \ket{0},
\end{equation}
where $\ket{0}$ represents the vacuum state, i.e.  $c_j\ket{0}=0$ for all $j$. This corresponds to filling the region $j<0$ with a gas of free fermions with density $\varrho=1$ that stochastically  expands at $t>0$ following the evolution of Eq.~\eqref{eq:model_dH}. The states $\ket{0}$ and $\prod_j c_j^{\dagger}\ket{0}$ are stationary under Eq.~\eqref{eq:model_dH}. However, when half of the chain is initialized in the former state and the second half in the latter state, as done in Eq.~\eqref{eq:DWdef}, a non-trivial out-of-equilibrium dynamics arises. As mentioned in the introduction, this protocol has been extensively studied when the hopping amplitudes are not random since it also provides the most simple setup to study transport of conserved charges~\cite{sk-23}.

\textbf{Free expansion (Sec.~\ref{sec:freexpsec}).} In this case, we consider a finite system of length $L$ with open boundary conditions. Initially, the system is prepared in the ground state of the Hamiltonian 
\begin{equation}
\label{eq:Ham_beta}
 H = -\frac{1}{2}\sum_{j=-L/2}^{L/2-1} (c^{\dagger}_{j+1} c_j + h.c.) + \sum_{j=-L/2}^{L/2} V_j(\beta) c^{\dagger}_j c_j,
\end{equation}
where $V_j(\beta) = \exp(\beta j)$ is a non-homogeneous confining potential that, in the limit $\beta \to \infty$, reproduces a hard-wall boundary condition at $j = 0$. That limit corresponds to a gas of free fermions at half-filling (density $\varrho=1/2$) confined in the region $j<0$ (the analysis can be straightforwardly extended away from half filling, without affecting any qualitative physical features). At $t=0$, the potential is released and the gas stochastically expands under Eq.~\eqref{eq:model_dH}. The ground state of Eq.~\eqref{eq:Ham_beta} differs from the domain wall state~\eqref{eq:DWdef} in several important respects. First, unlike the domain wall, it is not a product state and therefore exhibits a non-zero entanglement entropy already at $t=0$. Second, the particle density is $\varrho\leq1/2$ in the region $j<0$, which implies that only momentum modes with 
$k\in[-\pi/2,\pi/2]$ are occupied, whereas in the domain wall state all momentum modes are occupied in the region $j<0$.

In the following sections, we will study the time evolution of the entanglement entropy in these two protocols. To this end, we divide the system in the intervals $A = (-L/2,\ell]$ and $B=[\ell+1, L/2]$. The entanglement entropy of $A$ is given in terms of the reduced density matrix $\rho_{A}(t) = \text{Tr}_B[\rho_t]$ by
\begin{equation}\label{eq:vn_ent}
    S_A(t) = -\text{Tr}[\rho_A(t) \log \rho_A(t)].
\end{equation}
We will derive the exact time-dependent statistics of 
$S_A(t)$ at large space-time scales by applying methods from GHD to the stochastic dynamics governed by Eq.~\eqref{eq:model_dH}. We will check numerically these results using the fact that both initial states are Gaussian, i.e. they satisfy Wick theorem, and the Hamiltonian generator~\eqref{eq:model_dH} is quadratic in the fermionic operators $c_j$, $c_j^\dagger$. As a result, the time-evolved state~\eqref{eq:time_ev} remains Gaussian at all times~\cite{peschel-03, pe-09} and the entanglement entropy of each noise realization can be efficiently computed numerically for large systems by using the standard techniques described in Appendix~\ref{app:num_tech}.

\section{Diffusive domain wall melting }\label{sec:dwsec}

In this section, we study the quench from the domain-wall state in Eq.~\eqref{eq:DWdef}. Before focusing on this particular case, we briefly review the main tools used throughout the paper.
We are interested in describing the large-scale out-of-equilibrium properties of the system. Assuming that the system’s behavior varies slowly in space, then, under the scale-separation hypothesis~\cite{bcdf-16, cdy-16, dsvc-17}, it can be effectively divided into coarse-grained homogeneous cells comprising many lattice sites in the thermodynamic limit $L\to\infty$. Within this assumption, we adopt the well-known semiclassical approach based on the time evolution of the local Fermi occupation function~\cite{wigner-32, hpb-12}
\begin{equation}
\label{eq:Wigfundef}
    n_k(x,t) = \int dz \,\text{Tr}[ \rho_t \,c^{\dagger}_{x+z/2}\, c_{x-z/2}] \,e^{i kz },
\end{equation}
where we have replaced the site index $j$ by the continuum variable $x$.

In the case of deterministic, homogeneous, non-interacting dynamics, the Moyal equation governing the time evolution of $n_k(x,t)$ reduces, in the hydrodynamic limit $x,t\to\infty$ with fixed ratio $x/t$, to the Euler equation~\cite{bcdf-16, cdy-16, bfpc-18}
\begin{equation}
    \partial_t n_k(x,t) = -v_k \partial_x n_k(x,t),
\end{equation}
whose solution is given by 
\begin{equation}
\label{eq:euler_sol}
    n_k(x,t) = n_k(x-v_k t,0).
\end{equation}
According to this result, in the hydrodynamic limit, each mode $k$ remains uncorrelated with the others during the time evolution, due to the non-interacting nature of the dynamics, and propagates ballistically with velocity $v_k$. 

In our case, for the stochastic time evolution described by Eq.~\eqref{eq:model_dH}, we can consider the diffusive hydrodynamics in which each mode $k$ propagates with a  stochastic velocity $\xi_k(t)$ that depends on time.
Thus, in analogy to Eq.~\eqref{eq:euler_sol}, we expect the time evolution of $n_k(x,t)$ to satisfy
\begin{equation}
\label{eq:WigFunDyn}
n_k(x,t) = n_k(x-\xi_k(t),0).
\end{equation}
Since the Hamiltonian generator~\eqref{eq:model_dH} is quadratic
and the hopping amplitudes are spatially homogeneous, it is diagonal in momentum space,
\begin{equation}
dH_t=\sum_k d\varepsilon_{t, k} c_k^\dagger c_k,
\end{equation}
where $d\varepsilon_{t,k}=2\sqrt{D}{\rm Re}\{dW_t e^{ik}\}$ are real It\^o increments satisfying $d\varepsilon_{t, k}d\varepsilon_{t, q}=2Ddt\cos(k-q)$. Notice that the infinitesimal generators commute at different times. As a consequence, one can treat excitations with momentum $k$ as having a stochastic energy 
$\varepsilon_k(t)=2\sqrt{D}{\rm Re}\{W_t e^{ik}\}$, where $W_t$ is given by the sum of the It\^o increments $dW_t$ associated with the generator of the dynamics~\eqref{eq:model_dH}. Thus, we can identify $\xi_k(t) = \partial_k \varepsilon_k(t)= 2 \sqrt{D} \,\text{Re}\{W_t e^{ik}\}$ (up to an overall $\pi/2$ phase, which is irrelevant since the probability distribution of $W_t$ is invariant under multiplication by a phase). 
The dynamics of $n_k(x, t)$ is now not described by the Euler equation, but by the following stochastic differential equation in the It\^o sense
\begin{equation}
\label{eq:SDE_wigner}
    d n_k(x,t) =  D \, \partial_x^2 n_k(x,t) \,dt - \partial_x n_k(x,t) \, d\xi_k.
\end{equation}
This result follows from Eq.~\eqref{eq:WigFunDyn} by using the fact that $d\xi_k = 2\sqrt{D}(\cos k \,dB^1_t + \sin k \,dB_t^2)$, with $dB_t^{(1,2)}$ real It\^o increments satisfying $dB_t^2 = dt/2$.
In App.~\ref{app:deriv} we exactly derive Eq.~\eqref{eq:SDE_wigner} through a second-order gradient expansion of the Wigner function~\eqref{eq:Wigfundef}.

If we take the average over noise in Eq.~\eqref{eq:SDE_wigner}, we find the diffusion equation
\begin{equation}
\label{eq:diffeq}
    \partial_t \langle n_k(x,t) \rangle = D \, \partial_x^2 \langle n_k(x,t) \rangle.
\end{equation}
This discussion is valid for both the  quench protocols 
that we study in this paper.

\begin{figure*}[t]
\centering
\includegraphics[width=0.99\textwidth]{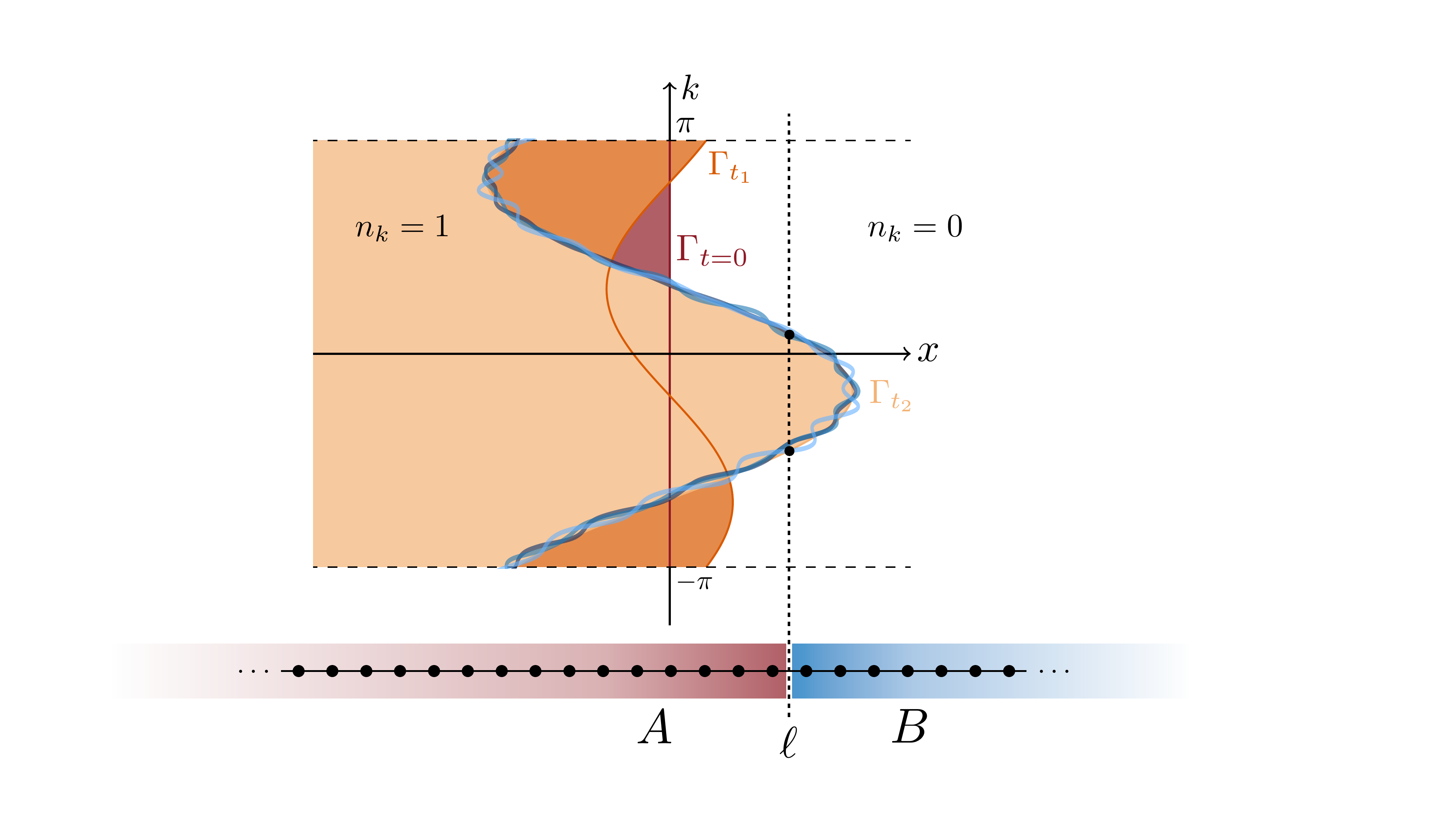}
    \caption{\label{fig:DQGHDsketch} Sketch of the time evolution of the local occupation function $n_k(x, t)$
    after a domain-wall quench in the $\nu=1$ QSSEP.  The dynamics of $n_{k}(x,t)$ follows the stochastic differential equation~\eqref{eq:SDE_wigner}. Colored regions correspond to $n_k(x, t)=1$ for single noise realizations at different times. We compute the statistical properties of the out-of-equilibrium entanglement entropy $S_{\ell}(t)$ between two intervals $A$ and $B$, connected at $x = \ell$. The exact contribution to entanglement of each noise realization is obtained using QGHD. In this framework, the quantum fluctuations of $n_k(x, t)$ are introduced in the form of a free massless boson living on  the Fermi contour $\Gamma_{t}$ that separates the fully filled and fully empty regions in the $(x,k)$ plane.  The contribution to the entanglement entropy of $A$ of each noise realization is given at, e.g., time $t = t_2$ by the correlation function of twist fields inserted at the Fermi points: the intersection of the line $x=\ell$ with the Fermi contour $\Gamma_{t_2}$ (black dots).
  }
\end{figure*}
We can now focus on the domain wall quench. The initial state~\eqref{eq:DWdef} has the following local Fermi occupation function,
\begin{equation}
\label{eq:WigFunInitDW}
n_{k}(x,t = 0) =
\begin{cases}
1, \qquad x \leq 0 \,\, \text{and} \,\, k \in [-\pi,\pi],\\
0, \qquad \text{otherwise}.
\end{cases}
\end{equation}
Therefore, in the hydrodynamic limit, the domain wall state can be seen as a gas of fermions with momentum $k\in[-\pi, \pi]$ that fill the spatial region $x\leq0$. In terms of the local occupation function, the average density of  particles at time $t$ is given by
\begin{equation}
\varrho(x,t) \equiv\int \frac{dk}{2\pi} \langle n_k(x,t)\rangle.
\end{equation}
Initially, $\varrho(x,t =0) = \theta(-x)$, with $\theta(x)$ the Heaviside step function. By solving the diffusion equation~\eqref{eq:diffeq} for these initial conditions, we find that, in the hydrodynamic limit, the average density of particles is
\begin{equation}
\label{eq:hydro_prof}
    \varrho(x,t) = \frac{1}{2}+ \frac{1}{2}\text{erf}\left(\frac{-\zeta}{2\sqrt{D}}\right),
\end{equation}
where $\zeta=x/\sqrt{t}$ and $\text{erf}(z)$ is the error function. This prediction can be verified by an exact microscopic computation of the average fermionic occupation function on the lattice, $\varrho_j(t) = \langle \text{Tr}[\rho_t\, c^{\dagger}_j c_j]\rangle$. 
Let us assume, for the moment, that the lattice has a finite length $L$ and periodic boundary conditions. Expressing the correlator in the momentum basis, we obtain
\begin{equation}
\varrho_j(t)= \sum_{j'} \mathcal{G}_{j-j'}(t) \varrho_{j'}(0),
\end{equation}
where
\begin{equation}
 \mathcal{G}_{j-j'}(t)=\frac{1}{L^2}\sum_{k, q}e^{i(k-q)(j-j')}\langle e^{i(\varepsilon_k(t)-\varepsilon_q(t))}\rangle.
 \end{equation}
Taking into account that $\langle e^{i(\varepsilon_k(t)-\varepsilon_q(t))}\rangle=e^{-4Dt\sin^2((k-q)/2)}$, we eventually find
\begin{equation}
 \mathcal{G}_{j-j'}(t)=\frac{1}{L}\sum_k e^{ik(j-j')} e^{-4Dt\sin^2(k/2)}.
\end{equation} 
In the thermodynamic limit $L \to \infty$, we have 
\begin{equation}\label{eq:mathcalG_thermo}
\mathcal{G}_{j-j'}(t) = e^{-2 Dt} I_{j-j'}(2 D t),
\end{equation}
where $I_x(z)$ is the modified Bessel function of first kind. If we take the diffusive hydrodynamic limit, i.e. $j-j' \to \infty$ and $t \to \infty$ with the ratio $(j-j')/\sqrt{t}$ fixed, in Eq.~\eqref{eq:mathcalG_thermo}, we obtain $\mathcal{G}_{j-j'}(t) \simeq e^{-(j-j')^2/(4 Dt )}/\sqrt{4\pi D t}$. This is precisely the propagator of the diffusion equation~\eqref{eq:diffeq}, whose convolution with $\varrho(x,0)$ provides the density profile in Eq.~\eqref{eq:hydro_prof}.
We remark that this diffusive transport is a result of averaging of different ballistic realizations. Since the diffusion constant $D$ only amounts to a rescaling of time in this model, we will set $D = 1$ in what follows.

\textbf{Entanglement spreading.} Let us now move on to study the time evolution of the entanglement entropy~\eqref{eq:vn_ent}. In principle, in the hydrodynamic limit, its leading contribution comes from the Yang–Yang entropy of the macrostate defined by the local occupation density $n_k(x)$ for the spatial region $A$~\cite{bfpc-18}
\begin{equation}
\label{eq:SYYdef}
    S_{{\rm YY}} = - \int_{x \in A} dx  \int \frac{dk}{2\pi}  \left[n_k(x)\log n_k(x) + (1- n_k(x)) \log(1-n_k(x))\right].
\end{equation}
However, in the protocols studied here, since for every noise realization the local occupation function $n_k(x, t)$ remains either 0 or 1 under the dynamics in Eq.~\eqref{eq:WigFunDyn}, the Yang-Yang entropy~\eqref{eq:SYYdef} is identically zero at all times. Thus to describe the spreading of the entanglement entropy in these protocols, we need to go beyond the semiclassical approach and reintroduce quantum fluctuations on top of the Euler hydrodynamics governing the dynamics of $n_k(x,t)$. This is the main result of Quantum Generalized Hydrodynamics (QGHD)~\cite{rcdd-21}. 

According to this approach, the quantum fluctuations of $n_k(x, t)$ at the boundary $\Gamma_t$ between the regions of the plane $(x, k)$ where $n_k(x,t) = 0$ and $n_k(x,t) = 1$ are described at leading order by a massless bosonic field whose dynamics is governed by an inhomogeneous Luttinger liquid Hamiltonian defined on $\Gamma_t$. 
This boundary $\Gamma_t$ is usually called Fermi contour.
To compute the entanglement entropy~\eqref{eq:vn_ent}, we consider the Rényi-$n$ entropy 
\begin{equation}
    S^{(n)}_A = \frac{1}{1-n} \log \text{Tr}[\rho_A^n],
\end{equation}
such that the von Neumann entropy~\eqref{eq:vn_ent} is obtained as $\lim_{n\to1} S_n(\rho_A) = S(\rho_A)$. As usual, for a subsystem $A=(-\infty, \ell]$ and integer values of $n$, $\text{Tr}[\rho_A^n]$ can be expressed as the one-point function of the twist field $\mathcal{T}_n(x)$ in the $n-$fold replicated theory~\cite{cc-04},
\begin{equation}
\label{eq:twist}
S_{\ell}^{(n)} = \frac{1}{1-n} \log \langle\mathcal{T}_n(\ell,t) \rangle.
\end{equation}
In the QGHD framework, this one-point function can be rewritten as a multi-point function of the chiral twist fields $\tau_n$, $\tilde{\tau}_n$ inserted at the Fermi points, i.e. the points of the Fermi contour $\Gamma_t$ that intersect the line $x=\ell$~\cite{scd-22}. These chiral twist fields are primary fields with conformal dimension $h_n = c/24\left(n - 1/n \right)$~\cite{cc-04}, where $c=1$ is the central charge of the Luttinger liquid theory. For simplicity, let us assume for the moment the presence of only two Fermi points. In the free expansion, the multiple–Fermi-point scenario will also appear. If we parametrize by $s$ the Fermi contour as $\Gamma_t=\{(x, k)=(x(s), k(s))\}$ and $s_{\pm}$ denote the position of the two Fermi points at time $t$ for a certain cut $x=\ell$, the one-point function on the right-hand side of Eq.~\eqref{eq:twist} is then expressed as the two-point function of the chiral twist fields $\tau_n,\tilde{\tau}_n$,
\begin{equation}
\label{eq:qghd_general}
\langle\mathcal{T}_n(\ell,t) \rangle = \varepsilon_n(\ell,t)^{2 h_n} \left| \frac{dx}{ds}\right|^{-h_n}_{s_+} \left| \frac{dx}{ds}\right|^{-h_n}_{s_-}   \langle \tau_n(s_+) \tilde{\tau}_n(s_-)\rangle.
\end{equation}
The $\varepsilon_n$ prefactor is a UV cut-off, which is non-universal in the sense that it depends on the specific lattice model. In our case, it is possible to determine it using the Fisher-Hartwig conjecture~\cite{jk-04, ce-10, aefs-14}, as explained below. For later convenience, we write the final result as 
\begin{equation}
\begin{split}
\label{eq:qghd_gen_fin}
    S^{(n)}_{\ell} &= \tilde{S}^{(n)}_{\ell} + \frac{2 h_n}{1-n} \log \varepsilon_n(\ell,t) \\
    &\equiv \tilde{S}_{\ell}^{(n)} + C_{\ell}^{(n)},
\end{split}
\end{equation}
where we have split the universal CFT prediction $\tilde{S}_{\ell}^{(n)}$ and the non-universal part due to the UV cut-off. We have also dropped the dependence on time for clarity.

We can apply these results from QGHD to obtain the 
statistics of the entanglement entropy under the 
stochastic unitary dynamics generated by 
Eq.~\eqref{eq:model_dH}. The main result of our work is that this can be achieved by averaging the QGHD prediction in Eq.~\eqref{eq:qghd_gen_fin} over the ensemble of Fermi contours $\{\Gamma_t\}$ corresponding to the different local occupation functions $n_k(x,t)$ generated by the stochastic differential equations~\eqref{eq:WigFunDyn}-\eqref{eq:SDE_wigner}, see Fig.~\ref{fig:DQGHDsketch} for a sketch.
Therefore, in the hydrodynamic limit, the average over all noise realizations of the R\'enyi entropy $S_{\ell}^{(n)}$ is given by its average over the Fermi contours $\{\Gamma_t\}$, 
\begin{equation}
\label{eq:exvcontour}
    \langle S_{\ell}^{(n)}\rangle \simeq \mathbb{E}_{\Gamma_t}[S_{\ell}^{(n)}],
\end{equation}
or, equivalently, over the corresponding Fermi points for the cut $\ell$.
From Eq.~\eqref{eq:WigFunDyn}, we can exactly determine the probability distribution of the possible Fermi contours $\Gamma_t$ at any time. To this end, it is useful to parametrize the different noise realizations of $\Gamma_t$ in the following way. In general, the Fermi contour is specified by a set of points in the $(x,k)$ plane. At $t = 0$:
\begin{equation}\label{eq:fermi_cont_0}
\Gamma_0=\{(x, k)=(x_0(k), k)\,|\, k\in\mathcal{I}_k\},
\end{equation}
where $\mathcal{I}_k=\{k\in[-\pi, \pi]\,|\, n_k(x_0(k), 0)=1\}$.
For the domain-wall state, $x_0(k) = 0$ for any $k$ and $\mathcal{I}_k=[-\pi, \pi]$, cf. Eq.~\eqref{eq:WigFunInitDW}. At time $t$, under Eq.~\eqref{eq:WigFunDyn}, the spatial points $x_0(k)$ are mapped to $x_0(k)+\xi_k(t)$. Since the Brownian motion $W_t$ that enters $\xi_k(t)$ is complex, we can write the later in the form $\xi_k(t)=2\rho_t\sin(k+\varphi_t)$, where $\rho_t\in[0, \infty)$ and $\varphi_t\in [-\pi, \pi]$ are independent random variables distributed according to
\begin{equation}\label{eq:prob_rho_phi}
p_\rho(\rho;t)=\frac{2\rho}{t}e^{-\rho^2/t}, 
\quad p_\varphi(\varphi; t)=\frac{1}{2\pi},
\end{equation}
respectively. Therefore, the Fermi contour of a single noise realization at time $t$ is
\begin{equation}\label{eq:Gamma_par}
 \Gamma_t=\{(x, k)=(x_0(k)+2\rho\sin(k+\varphi), k)\, | \, k\in\mathcal{I}_k\}.
\end{equation}
The time-dependent probability distribution of this Fermi contour factorizes in the probability distributions of $\rho_t$ and $\varphi_t$,
\begin{equation}
\label{eq:pdf_contour}
    p(\Gamma;t) \,d\Gamma = p_{\rho}(\rho;t) \,p_{\varphi}(\varphi;t) d\rho d\varphi.
\end{equation}

We can now analyze the average entanglement growth in the domain-wall quench. The strategy is the following: we determine the QGHD prediction in Eq.~\eqref{eq:qghd_gen_fin} for a Fermi contour with fixed parameters $\rho$ and $\varphi$, and then we calculate the average over all the possible Fermi contours~\eqref{eq:exvcontour} using the probability distribution~\eqref{eq:pdf_contour}. In the domain-wall case, the Fermi contour is periodic in $k$, so the random phase $\varphi$ can be dropped and the noise realizations are specified only by $\rho$. 
We can parametrize the Fermi contour for a given $\rho$ in terms of the coordinate $s$ as $k(s)=s$ and $x(s)=2\rho \sin(s)$. Then, for a given cut $x=\ell$ of the system, the two Fermi points are $s_+ =\pi - \arcsin(\ell/(2 \rho))$ and $s_-= \arcsin(\ell/(2 \rho))$. With this choice, since $s$ is a compact variable defined on a circle of length $2\pi$, we have~\cite{cc-04} 
\begin{equation}
    \langle \tau_n(s_+) \tilde{\tau}_n(s_-)\rangle = \left|2 \sin\left(\frac{s_+-s_-}{2}\right) \right|^{-2 h_n}.
\end{equation}
Therefore, the universal contribution to the von Neumann entropy $\tilde{S}^{(1)}_{\ell}$ in Eq.~\eqref{eq:qghd_gen_fin} is
\begin{equation}
    \tilde{S}^{(1)}_{\ell} = \frac{1}{6} \log \left[ 4 \rho \left(1-\frac{\ell^2}{(2 \rho)^2}\right)\right],\quad \text{if }\ell < 2\rho,
\end{equation}
and zero otherwise. As in the non-random case, the non-universal contribution in Eq.~\eqref{eq:qghd_gen_fin} is
given by the $\ell$-independent term of the entanglement entropy of $A = (-\infty,\ell)$ in a free-fermion state with a Fermi sea delimited by $s_\pm$. This term has been obtained using the Fisher-Hartwig conjecture in Refs.~\cite{jk-04, ce-10, aefs-14}. For conciseness, we directly report
its expression in the limit $n\to 1$, although it can be straightforwardly generalized to a generic Rényi index $n$,
\begin{equation}\label{eq:C_2_fp}
C_\ell^{(1)}=\frac{1}{2}\left(\Upsilon+\frac{\log2}{3}\right)+\frac{1}{6}\log\left|\sin\left(\frac{s_+-s_-}{2}\right)\right|, 
\end{equation}
where $\Upsilon$ is the constant $\Upsilon = 0.495018\dots$. Plugging in this formula the exact expression of the Fermi points $s_\pm$, we find
\begin{equation}
    C^{(1)}_{\ell} = \frac{1}{2}\left(\Upsilon + \frac{\log 2}{3}\right) + \frac{1}{12} \log  \left( 1- \frac{\ell^2}{(2 \rho)^2} \right).
\end{equation}
 Putting all the pieces together and after some straightforward algebra, we find that the contribution of a single noise realization is
\begin{equation}
\label{eq:qghd_rho}
    S_{\ell}(\rho)= \frac{1}{6}\log\left[2\rho \left(1-\left(\ell/2 \rho\right)^2 \right)^{3/2} \right] + \kappa_1, \quad \text{if }\ell < 2\rho 
\end{equation}
and $0$ otherwise. The constant term is $\kappa_1 = 0.478558\dots$. This result is identical to that obtained in the domain-wall melting setup with non-random hopping amplitudes~\cite{dsvc-17}, upon the substitution $2 \rho \to t$. 

\begin{figure*}[t]
\centering
\includegraphics[width=0.99\textwidth]{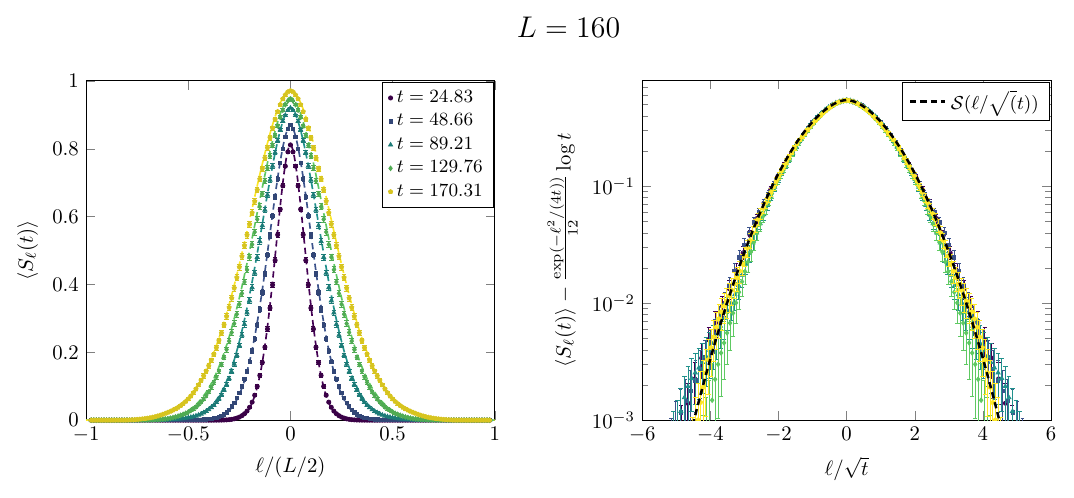}
    \caption{\label{fig:Sprof} Left panel: Profile of the average entanglement entropy for the subsystem $A = [-L/2,\ell]$ at different times $t$ starting from the domain-wall state~\eqref{eq:DWdef} in the $\nu=1$ QSSEP. The symbols are the exact average entanglement entropy over $\sim 10^3$ noise realizations, computed numerically in a lattice of size $L=160$. The errorbar is estimated from the standard deviation of the mean. The dashed curves correspond to the analytical prediction from QGHD in Eq.~\eqref{eq:avgSx0int}. Right panel: Same as in the left panel, but the symbols correspond to the exact numerical results after subtraction of the first term in Eq.~\eqref{eq:avgSx0int}, in order to extract the scaling function $\mathcal{S}(\ell/\sqrt{t})$. The data at different times all collapse onto our prediction for $\mathcal{S}(\ell/\sqrt{t})$ in Eq.~\eqref{eq:scal_fun_S} (black dashed line), clearly indicating diffusive spreading of the entanglement. 
  }
\end{figure*}

Finally, we average Eq.~\eqref{eq:qghd_rho} over the Fermi contours with respect to the probability distribution~\eqref{eq:pdf_contour} to derive the average entanglement entropy $\langle S_{\ell}(t) \rangle$ in the hydrodynamic limit,
\begin{equation}
\label{eq:avgSx0}
    \langle S_{\ell}(t)\rangle \simeq \int_{\rho>\ell/2}^{\infty} d\rho \, p_\rho(\rho; t) S_{\ell}(\rho).
\end{equation}
This integral can be computed analytically. After some algebra, we eventually obtain
\begin{equation}
\label{eq:avgSx0int}
    \langle S_{\ell}(t) \rangle \simeq \frac{e^{-\ell^2/(4 t)}}{12} \log(t)  + \mathcal{S}(\ell/\sqrt{t}),
\end{equation}
where 
\begin{equation}
\label{eq:scal_fun_S}
    \mathcal{S}(x) = \frac{e^{-x^2/4}}{12}\left( 3\log 4 - 3 \gamma_E + 12 \kappa_1 -2 \log x - 2 e^{x^2/4} \Gamma_0(x^2/4) \right).
\end{equation}
In this expression, $\gamma_{\rm E}$ is the Euler-Mascheroni constant and $\Gamma_0(x)$ stands for the incomplete Gamma function, $\Gamma_0(x) = \int_x^{\infty} dt \,e^{-t}/t$.
In Fig.~\ref{fig:Sprof} right, we plot the profile of the average entanglement entropy predicted by Eq.~\eqref{eq:avgSx0int} (solid lines) as a function of the subsystem cut $\ell$ at different times. We compare it with the exact values computed numerically in the lattice model (symbols), see Appendix~\ref{app:num_tech}, finding perfect agreement. In Fig.~\ref{fig:Sprof} right, we study the diffusive scaling behavior $\ell/\sqrt{t}$ predicted by the second term in Eq.~\eqref{eq:scal_fun_S}. We subtract the 
term $e^{-\ell^2/(4t)}/12\log(t)$ from the numerical data of Fig.~\ref{fig:Sprof} left, observing their collapse onto the curve in Eq.~\eqref{eq:scal_fun_S}.

\begin{figure*}[t]
\centering
\includegraphics[width=0.65\textwidth]{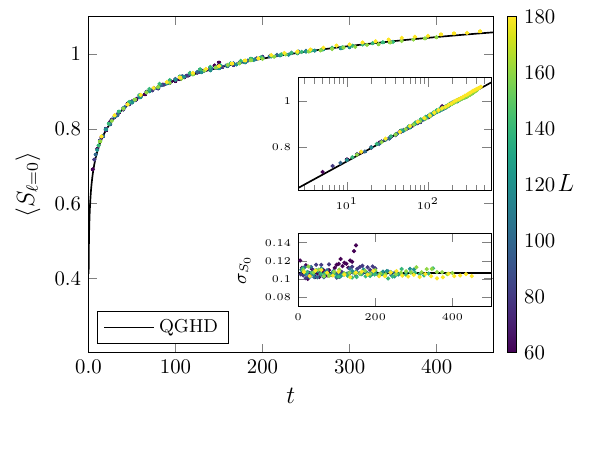}
    \caption{\label{fig:halfcutS} Time evolution of the average half-system entanglement in the $\nu=1$ QSSEP~\eqref{eq:model_dH} starting from the domain wall state \eqref{eq:DWdef}. The symbols correspond to the exact average value computed over $\sim 10^3$ noise realizations for increasing total system size $L$. The errorbar is estimated from the standard deviation of the mean. The continuous black curve corresponds to the analytical prediction in Eq.~\eqref{eq:halfsysS} derived using QGHD methods. The top inset shows the same data with logarithmic $x$-axis for a better visualization of the logarithmic growth of the average entanglement entropy with time, expected from Eq.~\eqref{eq:halfsysS}. The bottom inset shows the time dependence of the standard deviation $\sigma_S$ of the half-system entanglement entropy. The exact numerical data (symbols) tend to the QGHD prediction in Eq.~\eqref{eq:rel_fluct}, confirming that the   relative fluctuations of entanglement entropy vanish as $t\to \infty$.
  }
\end{figure*}

For the half-system entanglement entropy, i.e. $\ell=0$, the formula in Eq.~\eqref{eq:avgSx0int} simplifies considerably, 
\begin{equation}
\label{eq:halfsysS}
    \langle S_{\ell=0}(t) \rangle \simeq \frac{1}{12}\log t + C,
\end{equation}
with $C = \kappa_1 + (\log 4- \gamma_{\rm E})/12$. Compared to the quench with non-random hopping amplitudes, for which $S_0(t) \sim \frac{1}{6}\log t$~\cite{ep-14,dsvc-17}, the effect of stochastic hopping is to reduce the coefficient of the logarithmic term by a factor of $1/2$. This behavior is consistent with a diffusive scaling argument, whereby the ballistic time variable $t$ is effectively replaced by $\sqrt{t}$. In Fig.~\ref{fig:halfcutS}, we verify the prediction in Eq.~\eqref{eq:halfsysS} (solid line) against exact numerical results in the lattice model (symbols), showing a perfect agreement. 

Our approach also allows us to evaluate the out-of-equilibrium fluctuations of entanglement in the hydrodynamic regime. Specifically, we can obtain the variance of the entanglement entropy, $\sigma_S^2(t) = \langle S_{\ell}(t)^2\rangle - \langle S_{\ell}(t) \rangle^2$. As for the average entanglement entropy in Eq.~\eqref{eq:exvcontour}, we can compute $\langle S_{\ell}(t)^2\rangle$  from the QGHD prediction~\eqref{eq:qghd_rho} by averaging its square over all possible Fermi contours~\eqref{eq:Gamma_par}; that is,
\begin{equation}\label{eq:av_S2}
\langle S_\ell(t)^2\rangle\simeq \int_{\rho>\ell/2}^\infty d\rho\, p_\rho(\rho;t)S_\ell(\rho)^2.
\end{equation}
Combining this result with Eq.~\eqref{eq:avgSx0int}, we find that at large times
\begin{equation}
\label{eq:rel_fluct}
    \frac{\sigma_S(t)}{\langle S_{\ell}(t)\rangle} \underset{\substack{t\to\infty \\ \zeta = \ell/\sqrt{t} \text{ finite}}}{\simeq}  \sqrt{e^{\zeta^2/4}-1} + O\left(\frac{1}{\log t}\right).
\end{equation}
For half-system entanglement, i.e. $\zeta = 0$, the relative entanglement fluctuations vanish at long times. 
This means that the entanglement entropy is self-averaging in the diffusive hydrodynamic limit and, consequently, Eq.~\eqref{eq:avgSx0int} also corresponds to its typical value in that regime.
In the inset of Fig.~\ref{fig:halfcutS}, we check this result against exact numerical computations in a lattice of finite size $L$. As $L$ increases, the numerics (symbols) tend to the QGHD prediction for $\sigma_S$ (solid line).

\section{Free expansion}\label{sec:freexpsec}

\begin{figure*}[t]
\centering
\includegraphics[width=0.99\textwidth]{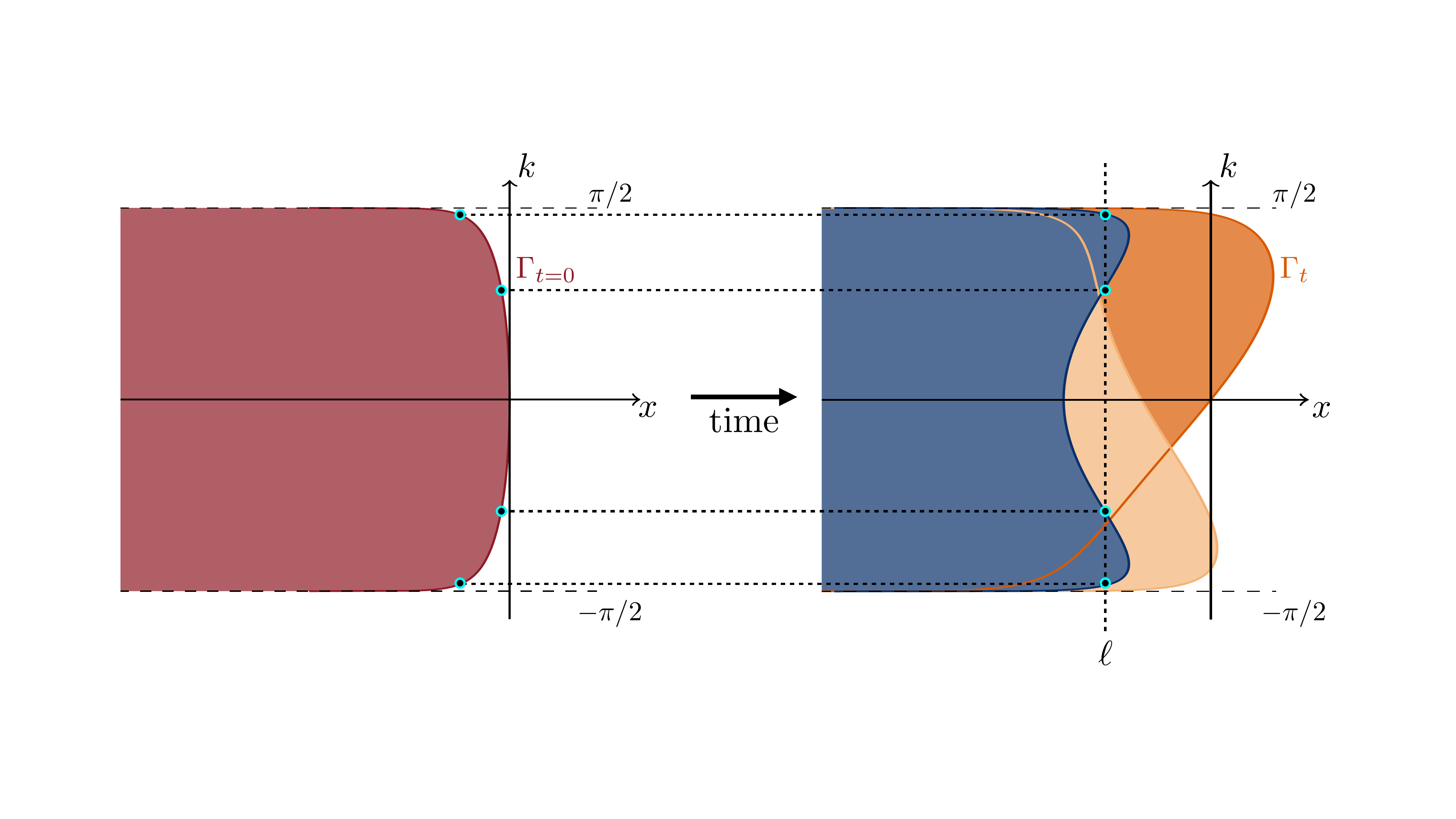}
    \caption{ Sketch of the time evolution
    of the local occupation function $n_k(x,t)$ in the free-expansion protocol studied in Sec.~\ref{sec:freexpsec}. The colored regions correspond to $n_k(x,t) = 1$. 
    In the left panel, we show the occupation function of the initial state, corresponding to the ground state of Eq.~\eqref{eq:Ham_beta} at finite $\beta$. 
    The Fermi contour $\Gamma_t$ evolves stochastically in time according to Eq.~\eqref{eq:SDE_wigner}. In the right panel, we show it for several noise realizations at time $t$. The average entanglement entropy $S_{\ell}(t)$ for $A = (-L/2,\ell]$ is obtained by averaging over the contribution of all possible Fermi contours. Each contribution is determined by the Fermi points: the intersections of the system cut $x=\ell$ with $\Gamma_t$. In the right panel, we highlight a noise realization with $4$ Fermi points, a scenario that does not occur in the domain-wall melting.
  }\label{fig:fexpQGHDsketch}
\end{figure*}

In this section, we discuss the second quench protocol, namely the free-expansion setup. We closely follow the approach of Ref.~\cite{skcd-21}, which considered a post-quench Hamiltonian with non-random hopping amplitudes. In this case, we have a finite system of length $L$ and open boundary conditions. The initial state is the ground state of the Hamiltonian~\eqref{eq:Ham_beta}.  Thus initially we have a gas of free fermions confined by the potential $V_j(\beta)$. 
For finite $\beta$, the local occupation function $n_k(x, 0)$ that describes this state in the hydrodynamic limit is 
\begin{equation}\label{eq:nkx_free}
n_k(x, 0)=\left\{\begin{array}{ll}
1, & |k|\leq \arccos V(x, \beta),\\
0, & \text{otherwise},
\end{array}\right.
\end{equation}
where $V(x, \beta)=\exp(\beta x)$. Essentially, the parameter $\beta$ acts as a regulator, 
ensuring that the initial Fermi contour $\Gamma_{0}$ is a smooth curve, which is required to apply QGHD. In Fig.~\ref{fig:fexpQGHDsketch}, we plot the occupation function~\eqref{eq:nkx_free} for a finite $\beta$.  The initial Fermi contour~\eqref{eq:Gamma_par} is then given by
\begin{equation}\label{eq:x_0_free}
    x_0(k) = \frac{1}{\beta} \log \cos k, \quad k \in[-\pi/2,\pi/2].
\end{equation}

Integrating the local occupation function~\eqref{eq:nkx_free} over the momenta, we obtain the initial density of particles
\begin{equation}
\label{eq:rhoinit_fe}
    \varrho (x, t= 0) = \int_{-\pi}^{\pi} \frac{dk}{2\pi} \, n_{k}(x,0) = \frac{1}{\pi}\arccos [\exp(\beta x)].
\end{equation}
In particular, $\varrho(x,t =0) = \theta(-x)/2$ in the hard-wall limit $\beta\to\infty$.
The noise-averaged density dynamics is then easily derived by solving the diffusion equation ~\eqref{eq:diffeq} with the initial condition in Eq.~\eqref{eq:rhoinit_fe}. This result is valid  in the space-time region that is not causally affected  by the hard-wall boundaries at $x=\pm L/2$. At these boundaries, the dynamics of the local occupation function is not solely described by Eq.~\eqref{eq:diffeq}, since one must also take into account the elastic reflection of the modes, as studied in Ref.~\cite{asw-22}. In our case, this would further require to replace the hard-wall boundaries at $x=\pm L/2$ by potentials similar to $V_j(\beta)$ that regularize the Fermi contour at these points.

\textbf{Entanglement spreading.} We move now on to  analyze the entanglement dynamics. To calculate the statistics of the entanglement entropy~\eqref{eq:vn_ent} in the hydrodynamic limit, we can follow the same strategy as in the domain-wall melting. As in that case, the $k$ modes propagate ballistically with stochastic velocity $\xi_k(t)$, and the Fermi contour is again of the form in Eq.~\eqref{eq:Gamma_par}, where the points $x_0(k)$ are now given by Eq.~\eqref{eq:x_0_free}; see Fig.~\ref{fig:fexpQGHDsketch} for a sketch of the time evolution of the Fermi contour in this setup.
 Each noise realization depends on the real parameters $\rho$ and $\varphi$ with time-dependent probability distributions in Eq.~\eqref{eq:prob_rho_phi}.
  We can then apply the QGHD framework introduced in the previous section to obtain the entanglement entropy of each noise realization and average over them to find the evolution of the average entanglement entropy. This approach only requires to determine the Fermi points at which the chiral twist fields $\tau_n$, $\tilde{\tau}_n$ must be inserted. For a subsystem $A=[-L/2, \ell)$ and fixed phase $\varphi$ and $\rho$, the Fermi points are the points in $\Gamma_t$ intersecting the system cut $x=\ell$; that is, the solutions of the equation
\begin{equation}
\label{eq:fermipts_eq}
    x_0(k) + 2 \rho \sin(k + \varphi) = \ell.
\end{equation}
Notice that, unlike in the domain-wall setting, only momenta $k\in[-\pi/2,\pi/2]$ are occupied in the initial state. As a consequence, the phase shift $\varphi$ cannot be dropped, as we did in the domain-wall case. This phase introduces a net imbalance between particles and holes crossing the cut at $x=\ell$, which has a non-trivial impact on the time evolution of the entanglement entropy for each realization of the noise.

In what follows, we found convenient to use  the following parametrization of the initial Fermi contour in terms of the real coordinate $s$ such that
\begin{equation}
\label{eq:param_s}
    \begin{cases}
        x_0(s) = -\frac{1}{\beta} \log \cosh(\beta s),  \\
        k(s) = -2 \arctan \tanh\left(\beta s/ 2\right).
    \end{cases}
\end{equation}
As shown in Ref.~\cite{skcd-21}, since the confining potential $V_j(\beta)$ at $t=0$ is not homogeneous, the Hamiltonian that governs the quantum fluctuations of $n_k(x, t)$ is that of a non-homogeneous Luttinger liquid with a spatially varying Fermi velocity $\sin( k(x, t))$. Given that $d x_0(s)/ds = \sin k(s)$, the choice in Eq.~\eqref{eq:param_s} removes this spatial dependence and guarantees that the two-point function of the chiral twists field $\tau_n$, $\tilde{\tau}_n$ on the initial Fermi contour is 
\begin{equation}
    \langle \tau_n(s_1) \tilde{\tau}_n(s_2) \rangle = (s_1-s_2)^{-2 h_n}.
\end{equation}

To properly account for the finite size of the system, we map the initial Fermi contour onto a circle by identifying $s_1 + L_s \sim s_1$, where $L_s = 2 (s(-L/2)-s(0))$ is the length of the circle, and $s(x_0)$ is the inverse function of $x_0(s)$ defined in Eq.~\eqref{eq:param_s}.
This is achieved by the conformal mapping
\begin{equation}
    s \mapsto \exp\left(2 i \pi \frac{s}{L_s}\right).
\end{equation}
The two-point function of twist fields on circle is then given by
\begin{equation}
\label{eq:2ptf_Ls}
    \langle \tau_n(s_1) \tilde{\tau}_n(s_2) \rangle = \left|\frac{L_s}{\pi} \sin\left(\frac{\pi}{L_s}(s_1-s_2) \right) \right|^{-2h_n}.
\end{equation}
Notice that, for large system sizes, $L_s = L + 2/\beta\,\log 2+O(e^{-2\beta L})$. Therefore, up to $L^{-1}$ corrections, it is possible to consider $L_s \simeq L$ when $L\gg1$. In the following, to simplify the notation, we introduce the function
\begin{equation}
    g(u,v;L) \equiv \frac{L_s}{\pi} \,\left|\sin\left(\frac{\pi(u-v)}{L_s}\right)\right|.
\end{equation}

The prediction from QGHD for a single noise realization, with parameters $\rho$ and $\varphi$, depends on the number of Fermi points that solve Eq.~\eqref{eq:fermipts_eq}. If there are only two Fermi points, then the QGHD prediction is the same as in the previous section, see Eqs.~\eqref{eq:qghd_general} and~\eqref{eq:qghd_gen_fin}. Let $s_{\pm}$ be the two solutions of Eq.~\eqref{eq:fermipts_eq} in terms of the parametrization of the Fermi contour in Eq.~\eqref{eq:param_s}. The contribution to the von Neumann entanglement entropy of that noise realization is
\begin{multline}
\label{eq:S_2p}
    S_{\ell}(\rho,\varphi)|_{\mathrm{(2p)}} = \frac{1}{6} \log \left( \frac{L_s}{\pi} \left|\frac{dx}{ds} \right|^{1/2}_{s_+} \left|\frac{dx}{ds} \right|^{1/2}_{s_-} g(s_+,s_-;L) \right) \\
    + \frac{1}{6} \log \left|\sin \frac{k(s_+)-k(s_-)}{2}\right| + \frac{1}{2} \left(\Upsilon + \frac{\log 2}{3}\right).
\end{multline}
The last two terms in the latter expression correspond to the non-universal lattice contribution $C_\ell$, cf. Eq.~\eqref{eq:C_2_fp}.

Unlike the domain-wall melting, there can also be four Fermi points $s_i$, with $i=1,\dots, 4$,  solving Eq.~\eqref{eq:fermipts_eq}. Let $u_j = s_{2j-1}$ and $v_j = s_{2j}$, these points describe a split Fermi sea, i.e. $n_k(\ell, t)=1$ for $k\in[k(u_1), k(v_1)]\cup [k(u_2), k(v_2)]$ (see Fig.~\ref{fig:fexpQGHDsketch} for an explicit example of this scenario). The universal CFT contribution to the $n$-th Rényi entropy is given by
\begin{equation}
\tilde{S}^{(n)}_{\ell}(\rho,\varphi) |_{\mathrm{(4p)}} = \frac{1}{1-n}\log\left[ \left(\prod_{i=1}^2 \left|\frac{dx}{ds} \right|^{-h_n}_{u_i} \left|\frac{dx}{ds} \right|^{-h_n}_{v_i}\right) \langle \tau_n(u_1)\tilde{\tau}_n(v_1)\tau_n(u_2)\tilde{\tau}_n(v_2)\rangle\right],
\end{equation}
where the four-point function of the twist fields is~\cite{cc-04, cfh-05} 
\begin{equation}
    \langle \tau_n(u_1)\tilde{\tau}_n(v_1)\tau_n(u_2)\tilde{\tau}_n(v_2)\rangle = \left( \frac{g(u_1,u_2) g(v_1,v_2)}{g(u_1,v_1) g(u_1,v_2) g(u_2,v_1) g(u_2,v_2) }\right)^{2 h_n}.
\end{equation}
In the replica limit $n\to1$, we obtain
\begin{equation}
\label{eq:S_4p}
    \tilde{S}_{\ell}(\rho,\varphi)|_{\mathrm{(4p)}} = \frac{1}{6}\log \left[\left(\prod_{i=1}^2 \left|\frac{dx}{ds}\right|_{u_i}^{1/2}\left|\frac{dx}{ds}\right|_{v_i}^{1/2} \right) \frac{g(u_1,v_1) g(u_1,v_2) g(u_2,v_1) g(u_2,v_2)}{g(u_1,u_2) g(v_1,v_2)}\right].
\end{equation}
The explicit expression of the Jacobian is $dx/ds =- \mathrm{sech} (\beta s) \, [ \sinh(\beta s) + 2 \beta \rho \cos( k(s) + \varphi) ]$.

The non-universal lattice contribution is the same as the $\ell$-independent term of the entanglement entropy of the interval $A=[-L/2, \ell)$ in a free-fermion state with a split Fermi sea defined by the moments $[k(u_1), k(v_1)]\cup[k(u_2), k(v_2)]$. This term can be again derived by applying the Fisher-Hartwig conjecture, as shown in Ref.~\cite{aefs-14}. The result in the limit $n\to 1$ is
\begin{equation}
\label{eq:C14p}
    C^{(1)}_{\ell}|_{(4p)} = -\frac{1}{6} \sum_{1\leq i <j \leq 4} (-1)^{i+j} \log \left| \sin \left(\frac{k_{F,i}-k_{F,j}}{2} \right)\right| + \Upsilon + \frac{\log 2}{3},
\end{equation}
where $k_{F,i} \equiv k(s_i)$.
The total the entanglement entropy is then the sum of Eqs.~\eqref{eq:S_4p} and \eqref{eq:C14p}, $S_{\ell}(\rho, \varphi)|_{(4p)} = \tilde{S}_{\ell}(\rho, \varphi)|_{(4p)} + C^{(1)}_{\ell}|_{(4p)}$.

It is worth mentioning that, upon replacing $2\rho \mapsto t$ and fixing the 
phase $\varphi$, the Fermi contour $\Gamma_t$ coincides with that of a free expansion driven by the non-random Hamiltonian $H = \sum_{j} e^{i \varphi} c^{\dagger}_{j} c_{j+1} + h.c.$. As a consequence, the above QGHD predictions also apply to this case. This setting is analyzed in Appendix~\ref{app:freeexp}, where we provide a numerical verification of the results obtained here for a single realization.

\begin{figure*}[t]
\centering
\includegraphics[width=0.65\textwidth]{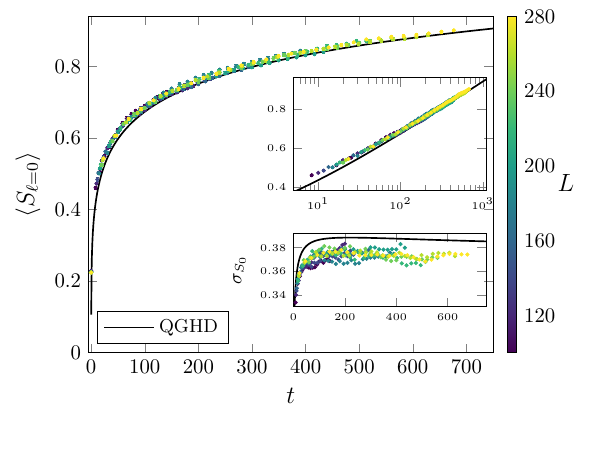}
    \caption{\label{fig:S0exp} Average half-system entanglement entropy $\langle S_{\ell = 0}(t) \rangle$ as a function of time in the free expansion protocol, starting from the ground state of the Hamiltonian~\eqref{eq:Ham_beta} with $\beta = 0.25$. The symbols are exact average entropy over~$\sim 10^3$ noise realizations. The corresponding errorbar is the standard deviation of the mean. We show results for increasing system sizes $L$, up to times in which the effects of boundaries at $x=\pm L/2$ are negligible.  The solid curve is the QGHD prediction. Top inset: Same data as in the left panel but in logarithmic time scale. After times of order $\sim 10$, the exact numerical results tend to the QGHD prediction. Bottom inset: time dependence of the standard deviation $\sigma_S$ of the half-system entanglement entropy. Symbols are the exact numerical data and the solid curve is the QGHD prediction.
  }
\end{figure*}

In summary, each noise realization corresponds to a Fermi contour $\Gamma_t$ at time $t$ characterized by the parameters $\rho$ and $\varphi$. For a bipartition of the system at $x=\ell$, the Fermi points are determined by solving numerically Eq.~\eqref{eq:fermipts_eq}. Once the solutions are obtained, the QGHD contribution of the noise realization is given by Eq.~\eqref{eq:S_2p} when there are two solutions, and by Eqs.~\eqref{eq:S_4p}–\eqref{eq:C14p} when there are four. The last step is averaging over all the possible Fermi contours with the probability distribution in Eq.~\eqref{eq:pdf_contour},
\begin{equation}
\label{eq:qghd_fe}
    \langle S_{\ell}(t) \rangle \simeq \int d\rho \,d\varphi \,\,p_{\rho}(\rho;t) \,p_{\varphi}(\varphi;t) \, S_{\ell}(\rho,\varphi),
\end{equation}
to obtain the evolution of the average entanglement entropy of that bipartition in the hydrodynamic limit. In the hard-wall limit $\beta \to \infty$, we find that the leading behaviour of the half-system entropy is
\begin{equation}
    \langle S_{\ell = 0}(t)\rangle = \frac{1}{8} \log t + \rm{cst.},
\end{equation}
see in Appendix~\ref{app:freeexp} a discussion of this limit. 
We can derive in the same way the higher moments $\langle S_\ell(t)^n\rangle$ of the entanglement entropy; in particular, its variance
$\sigma_S^2$, as we did in Eq.~\eqref{eq:av_S2} in the domain-wall melting. In the large $\beta$ limit and $\ell=0$, the relative entanglement fluctuations vanish at large times, i.e. $\sigma_S/\langle S_{\ell=0}(t)\rangle\to 0$ as $t\to\infty$, indicating that in this case $\langle S_{\ell = 0}(t)\rangle$ corresponds to the typical value of the entanglement entropy.

In Fig.~\ref{fig:S0exp}, we provide a first numerical check of the prediction in Eq.~\eqref{eq:qghd_fe} for an initial trap potential with finite $\beta$. We show the time evolution of the half-system average entanglement entropy ($\ell=0)$ for different total system sizes. The QGHD-based prediction (solid line) agrees well with the exact numerical results (symbols) for large enough times. In the lower inset, we show the standard deviation of the entanglement entropy, compared with the QGHD prediction (solid line). Although slowly decaying corrections to the hydrodynamic prediction are present, both exact numerics and QGHD tend to a constant value at late times, as in the $\beta\to\infty$ limit and the domain-wall melting.

\begin{figure}[t]
    \centering
    \includegraphics[width=.9\linewidth]{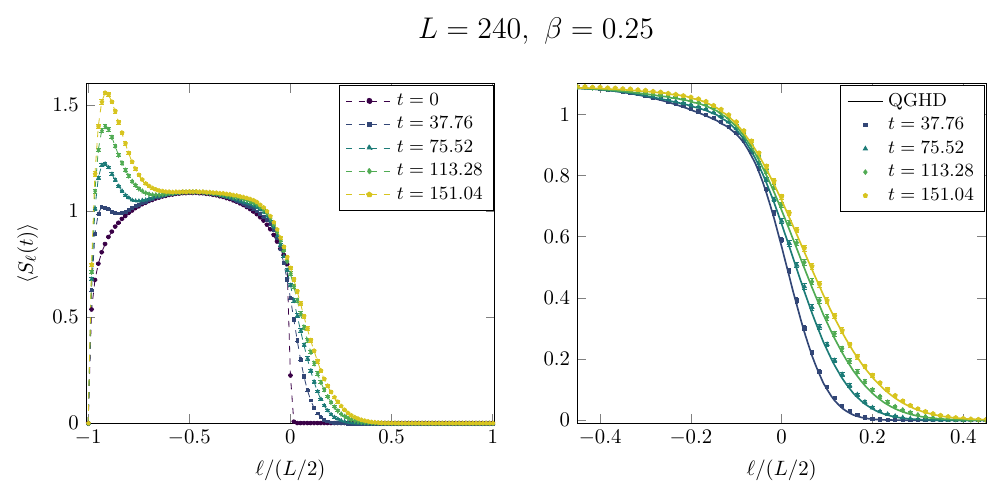}
\caption{\label{fig:Sprofexp} Left panel: Profile of the average entanglement entropy as a function of the system cut $\ell$ at different times in the free-expansion protocol. We take a system of length $L = 240$ and an initial confining potential with $\beta = 0.25$. The symbols represent the exact average entropy computed numerically by sampling 
$\sim 10^3$ noise realizations. The dashed curves are only meant to guide the eye. Notice that the entanglement entropy evolves not only near $\ell=0$, the separation between the initially filled and empty regions, but also near the system boundary $\ell=-L/2$. This boundary effect is not captured by our hydrodynamic framework, but could in principle be incorporated through an appropriate modification (see main text).  Right panel: Zoom of the region $\ell \in [-L/4,L/4]$, where the boundary effects are negligible for the times considered. The solid curves correspond to our QGHD prediction~\eqref{eq:qghd_fe}.}
\end{figure}

 In Fig.~\ref{fig:Sprofexp} left, we show the profile of the average entanglement entropy as we vary the system cut $\ell$ at different times. At $t=0$, the entanglement entropy is not zero in the region $j<0$ where the gas is confined. Observe that, after the potential is released and the system is driven by Eq.~\eqref{eq:model_dH}, the entanglement entropy evolves not only around the point $x=0$ but also near the system boundary at $x=-L/2$. This behavior is absent in the non-random case studied in Ref.~\cite{skcd-21}, where the phase $\varphi$ is equal to zero. A non-zero phase $\varphi$ causes particles elastically reflected at $x=-L/2$ to populate modes outside the interval $[-\pi/2,\pi/2]$, which in turn leads to a non-trivial evolution of the entanglement entropy. This effect may be captured within our QGHD framework by regularizing the Fermi contour $\Gamma_t$ at $x=-L/2$, replacing the hard wall with a smooth exponential potential, and explicitly accounting for the elastic reflection of modes at this point in the evolution of $\Gamma_t$. 
 The Fermi contour evolution considered in Eq.~\eqref{eq:Gamma_par} applies in the space-time region causally disconnected from the boundaries. As shown in the right panel of Fig.~\ref{fig:Sprofexp}, the QGHD prediction derived from this evolution correctly captures the spreading of entanglement for system cuts $\ell$ sufficiently far from $x=-L/2$ and for times when the influence of the boundaries is negligible.

\section{Conclusions}\label{sec:conclusions}

We have studied the unitary stochastic dynamics of a free-fermion system with time-dependent random hopping amplitudes that are spatially homogeneous, starting from spatially inhomogeneous initial states. In particular, we considered two paradigmatic initial configurations: a domain-wall state and a gas of free fermions initially confined to half of the system.
Owing to the spatial uniformity of the noise, the dynamics of the $\nu=1$ QSSEP admits a simple and intuitive description at large space–time scales in terms of quasiparticles undergoing Brownian motion. Leveraging this property, we adapted the tools of quantum generalized hydrodynamics \cite{rcdd-21}, originally developed for inhomogeneous quenches with deterministic hopping amplitudes, to the present stochastic setting. As in the deterministic case, the system can be locally characterized by a coarse-grained occupation function in phase space. We derived the stochastic evolution equation governing this occupation function and showed that, upon averaging over noise realizations, it obeys a diffusion equation, in contrast to the Euler-type equation that arises in the non-random case.
Finally, we derived the exact asymptotic behavior of entanglement spreading by explicitly incorporating quantum fluctuations of the occupation function and, using conformal field theory methods, deriving the contribution of each individual noise realization to the entanglement entropy. By averaging over all possible realizations of the noise, we obtained the exact full counting statistics of the entanglement entropy in the hydrodynamic regime. Our results provide a rare example of an exact and fully controlled characterization of entanglement growth in noisy quantum many-body systems, complementing recent exact results obtained for global quantum quenches~\cite{alba-25,raca-25}.

While our analysis has primarily focused on the entanglement entropy, the hydrodynamic framework developed here can be straightforwardly extended to investigate other observables, such as density–density correlation functions~\cite{dds-24}, symmetry resolved entanglement \cite{sh-22}, and entanglement asymmetry \cite{raca-25,amc-22}. 
It can also be readily applied to study other variants of the problems already analyzed in the non-random model, for instance multiple domain walls in the initial state~\cite{scd-22}, the presence of reflecting boundaries~\cite{asw-22}, or including a conformal defect at a given spatial point in the system~\cite{ep-12, fg-21, csrc-23} as well as the presence of a linear potential \cite{cvcg-23, ebs-25}. 
In addition, it would be natural to generalize the present framework to initial thermal states or to junctions of ground states corresponding to different free-fermion Hamiltonians, opening the way to a broader class of non-equilibrium protocols within the same hydrodynamic description.

Another interesting direction is to investigate the setups considered here in a generalization of the $\nu=1$ QSSEP with interacting terms, such as a homogeneous noisy version of the XXZ spin-$1/2$ chain \cite{bastianello20,wlzz-26}. In the gapless phase of this model, the exact profiles of conserved charges and currents have been derived by means of GHD in various inhomogeneous quenches~\cite{bcdf-16, clv-18, clcd-20}, including domain-wall melting. However, while the spreading of entanglement in this model has been studied and several analytic conjectures for its asymptotic behavior have been proposed~\cite{ah-14, clcd-20, scd-22}, a systematic approach based on requantizing the occupation function, analogous to the one applied here, is still missing. 
\newline

\textbf{Acknowledgments.} The authors thank Stefano Scopa for useful discussions. P.C., F.A., and A.R. acknowledge support from the European Research Council under the Advanced Grant no. 101199196 (MOSE). 
V.A. has been supported by the project ``Artificially devised many-body quantum dynamics in low dimensions - ManyQLowD'' funded by the MIUR Progetti di Ricerca di Rilevante Interesse Nazionale (PRIN) Bando 2022 - grant 2022R35ZBF. This article is based upon work from COST Action "Many-body Open Quantum Systems" (QOpen) CA24109 supported by COST (European Cooperation in Science and Technology).

\appendix

\section{Derivation of Eq.~\eqref{eq:SDE_wigner}}
\label{app:deriv}
We provide here an analytical derivation of Eq.~\eqref{eq:SDE_wigner}. Our starting point is the definition of the Wigner function \eqref{eq:Wigfundef}, which we rewrite here for clarity,
\begin{equation}
\label{eq:appwigfundef}
    n_k(x,t) = \int dz \,\text{Tr}[ \rho_t \,c^{\dagger}_{x+z/2}\, c_{x-z/2}] \,e^{i kz }.
\end{equation}
The time evolution of the two-point correlation matrix 
\begin{equation}
    G_{lm}(t) =  {\rm Tr} [\rho_tc_l^{\dagger}c_m],
\end{equation}
is given by 
\begin{equation}
\label{eq:sevolG}
G(t+dt) = e^{i  dh_{t}} \,G(t) \, e^{-i dh_t},
\end{equation}
with $(dh_t)_{ij} = dW_t \,\delta_{i,j+1} + \overline{d W_t}\delta_{i,j-1}$. We recall that the complex It\^o increment satisfies $dW_td W_t =\overline{dW}_t\overline{d W_t}= 0$ and $dW_t\overline{d W_t} = dt$ (we fix $D = 1$). Expanding the exponentials in Eq.~\eqref{eq:sevolG} and keeping terms up to order $dt$, we obtain the following stochastic differential equation for $G_{lm}$
\begin{multline}
\label{eq:appsdeG}
    d G_{lm} = dt (-2 G_{lm} + G_{l+1,m+1}+G_{l-1,m-1})\\- i dW_t (G_{l+1,m} - G_{l,m-1}) - i \overline{dW_t} (G_{l-1,m}-G_{l,m+1}).
\end{multline}
We now consider the center of mass $x$ and relative coordinates $z$ exactly as in Eq.~\eqref{eq:appwigfundef}, 
\begin{equation}
    \begin{cases}
        x = \frac{l+m}{2}, \\
        z = l - m,
    \end{cases}
\end{equation}
such that $G_{x,z} = \text{Tr}[\rho_t c^{\dagger}_{x+z/2} c_{x-z/2}]$.
In these coordinates
\begin{multline}
    d G_{x,z} = dt (-2 G_{x,z}+G_{x+1,z}+G_{x-1,z}) \\- i dW_t( G_{x+1/2,z+1}- G_{x-1/2,z+1}) - i \overline{dW_t} (G_{x-1/2,z-1}-G_{x+1/2,z-1}).
\end{multline}
We can perform in this expression a second-order gradient expansion by assuming $G_{x,z}$ to be slowly varying with respect to $x$. In this way, one finds, up to order $dt$
\begin{equation}
\label{eq:appSDEG}
    d G_{x,z} = dt\, \partial_x^2 G_{x,z}-i dW_t \,\partial_x G_{x,z+1} + i \overline{dW_t} \,\partial_x G_{x,z-1}.
\end{equation}
Taking the Fourier transform with respect to $z$ as in Eq.~\eqref{eq:appwigfundef}, we obtain
\begin{equation}
    d n_{k}(x,t) = \partial_x^2 n_{k}(x,t) \,dt - \partial_x n_k(x,t) \,2\,{\rm Re} \{ dW_t e^{ik}\},
\end{equation}
which exactly corresponds to Eq.~\eqref{eq:SDE_wigner} of the main text, after having absorbed the irrelevant phase $i$ into the complex It\^o increment $dW_t$ in Eq.~\eqref{eq:appSDEG}.

\section{Free expansion of a hard-core quantum gas with complex hopping}
\label{app:freeexp}

In this Appendix, we study entanglement spreading in the free expansion protocol considering a post-quench Hamiltonian with non-random, complex hopping amplitudes. In particular, the initial state is the same as in the main text, i.e. the ground state of the Hamiltonian~\eqref{eq:Ham_beta}, but the system now evolves under the deterministic unitary dynamics generated by the Hamiltonian
\begin{equation}
\label{eq:HXX_phi}
    H_{\varphi} = -\sum_{j=-L/2}^{L/2-1} e^{i \varphi} c^{\dagger}_j c_{j+1} + e^{-i\varphi} c^{\dagger}_{j+1} c_j.
\end{equation}
The case $\varphi=0$ was studied in Ref.~\cite{skcd-21}. As explained in the main text, extending the analysis to $\varphi \neq 0$ is essential for understanding the entanglement dynamics in the $\nu = 1$ QSSEP model. The evolution of the Fermi contour under this Hamiltonian coincides with that of a single noise realization in the $\nu = 1$ QSSEP, upon replacing the stochastic parameter $2\rho$ with time $t$. Thus, the entanglement statistics are obtained by properly averaging the different contributions arising from the dynamics generated by Eq.~\eqref{eq:HXX_phi}. Yet studying the evolution under the Hamiltonian~\eqref{eq:HXX_phi} for a fixed $\varphi \neq 0$ is interesting in its own right. The phase $\varphi$ shifts the quasiparticle velocity, $v_k = \sin(k+\varphi)$, thereby directly affecting the entanglement spreading.

\begin{figure*}[t]
\centering
\includegraphics[width=0.65\textwidth]{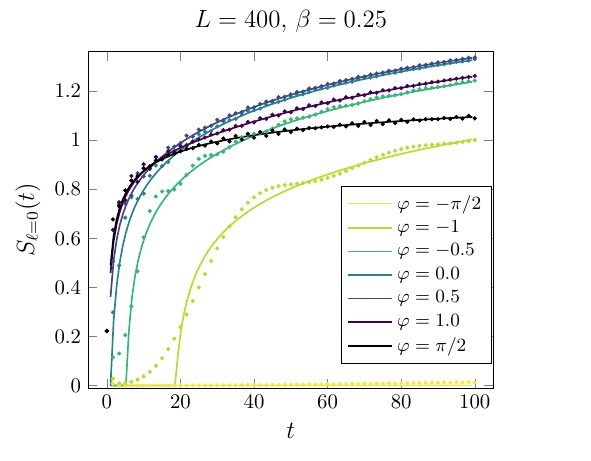}
    \caption{\label{fig:S0exp_xxphi} Half-system entanglement entropy $S_{\ell = 0}$ as a function of time in the free-expansion protocol with deterministic unitary dynamics given by $H_{\varphi}$ in Eq.~\eqref{eq:HXX_phi}. The parameter $\varphi$ controls the density current along the cut of the system.
    The symbols are the exact entanglement entropy computed numerically from the two-point correlation matrix. The solid lines correspond to the QGHD prediction, obtained by replacing $2\rho\to t$ and fixing the value of $\varphi$ in the results of Sec.~\ref{sec:freexpsec} for a single noise realization. Depending on the value of $\varphi$, the size of corrections to the hydrodynamic prediction at intermediate times can be significant.}
\end{figure*}

In Fig.~\ref{fig:S0exp_xxphi}, we show the time evolution of the half-system entanglement entropy for different values of $\varphi$. The solid curves correspond to the QGHD prediction, obtained by adapting the results in Sec.~\ref{sec:freexpsec} for a single noise realization in the $\nu = 1$ QSSEP. Overall, we observe good agreement with the exact numerical results (symbols), calculated using the methods in Appendix~\ref{app:num_tech}, which further confirms the validity of the predictions in Sec.~\ref{sec:freexpsec}. Notice that for some choices of the phase $\varphi$, larger oscillations in the entanglement entropy appear at intermediate times, which are not captured by the QGHD prediction. These corrections to the hydrodynamic prediction explain the deviation in the average entanglement in the $\nu = 1$ QSSEP shown in Fig.~\ref{fig:S0exp} for times $t \sim 10$, before the onset of hydrodynamic behavior.

A relevant special case is the hard-wall limit $\beta\to\infty$ of the confining potential at $t=0$. By solving Eq.~\eqref{eq:fermipts_eq} (with the substitution $2\rho \to t$) approximately up to order $\beta^{-1}$ and taking the infinite system size limit $L\to \infty$, we obtain for the half-system entanglement entropy 
\begin{equation}
\label{eq:Slargebeta}
S_{\ell =0}(t,\varphi) = \begin{cases}
\frac{1}{4}\log \left(t \cos \varphi \left(\frac{1+\sin\varphi}{2}\right)^{1/3} \right) + \rm{cst.}  \qquad \varphi \in (-\pi/2, \pi/2) , \\
0 \qquad \varphi=-\pi/2.
\end{cases}
\end{equation}
This result is consistent with that found in Ref.~\cite{skcd-21} for $\varphi=0$ and shows that the half-system entanglement entropy grows logarithmically, $S_{\ell = 0}(t) \sim \frac{1}{4} \log t + \rm{cst.}$, with a coefficient independent of the phase $\varphi$. Replacing $t \to 2 \rho$ in this asymptotic behavior and inserting it in Eq.~\eqref{eq:qghd_fe}, we obtain that the average half-system entropy grows as $\langle S_{\ell = 0} (t) \rangle \sim \frac{1}{8} \log t + \rm{cst.}$ in the $\nu = 1$ QSSEP free-expansion.

\section{Numerical techniques}
\label{app:num_tech}
In this Appendix, we describe the numerical method employed to check the results obtained 
in the main text. The initial states considered in this work are Gaussian and have a well-defined number of particles. Both properties are preserved under the time evolution~\eqref{eq:model_dH} of the $\nu=1$ QSSEP, and the time-evolved states are fully characterized by their two-point correlation matrix
\begin{equation}
    G_{ij}(t) =  {\rm Tr} [\rho_tc_i^{\dagger}c_j].
\end{equation}
If $G$ is known, its restriction $G_A$ to a subsystem $A$ provides the entanglement entropy~\eqref{eq:vn_ent} through the formula~\cite{peschel-03,pe-09}
\begin{equation}
    S_A = -\text{Tr}[G_A \log G_A + (\mathbb{I}-G_A) \log (\mathbb{I}-G_A)].
\end{equation}
We numerically compute the stochastic time evolution of $G(t)$ for a single noise realization as follows.
We consider a small time step $dt = 0.05$ and we check a posteriori that decreasing $dt$ does not affect considerably the results. For each time step, we evolve $G(t)$ as 
\begin{equation}
    G(t+dt) = e^{i  dh_{t}} \,G(t) \, e^{-i dh_t},
\end{equation}
where $(dh_t)_{ij} = dW_t \,\delta_{i,j+1} + \overline{d W_t}\delta_{i,j-1} $ is the single-particle Hamiltonian of the generator~\eqref{eq:model_dH}. The stochastic hopping amplitude $dW_t$ is randomly generated such that $dW_t = x_t + i y_t$ where $x_t,y_t$ are normally distributed variables with zero mean and $D dt /2$ variance. We then sample various different trajectories by repeating the same steps.

Finally, we need to specify the correlation matrix $G(t)$ at initial time $t =0$ for the two protocols studied. 
For the domain wall state defined in Eq.~\eqref{eq:DWdef},
\begin{equation}
    G_{\rm{DW}}(t = 0)_{ij} = \begin{cases}
        \delta_{ij} \qquad i, j \leq 0, \\
        0 \qquad\text{otherwise}.
    \end{cases}
\end{equation}
In the free expansion setup, $G(t=0)$ is the two-point correlation matrix of the ground state of Eq.~\eqref{eq:Ham_beta}. It can be obtained by diagonalizing the corresponding single-particle Hamiltonian
\begin{equation}
\label{eq:h1p}
    h_{ij} = -\frac{1}{2}(\delta_{i,j+1} + \delta_{i,j-1}) + V_i(\beta) \delta_{ij}.
\end{equation}
Let $\mathbb{U}$ be the matrix that diagonalizes $h$ in Eq.~\eqref{eq:h1p}, $(\mathbb{U}^T h \mathbb{U})_{ij} = \varepsilon_i \delta_{ij}$. The columns of $\mathbb{U}$ are the eigenvectors $\{\bm{u}^i \}_{i = 1,\dots,L}$ of $h$  with eigenvalues $\{\varepsilon_i\}_{i=1,\dots,L}$, i.e. $\mathbb{U}_{ij} = u_i^j$. The two-point correlation matrix of the ground state of Eq.~\eqref{eq:Ham_beta} is then obtained as 
\begin{equation}
    G(t = 0) = \mathbb{U}\, D\, \mathbb{U}^{T},
\end{equation}
where $D_{ij} = \delta_{ij}  \theta(-\varepsilon_i)$, and $\theta(x)$ is the Heaviside theta function.

\end{document}